\documentclass[journal]{IEEEtran}

\usepackage{cite}

\ifCLASSINFOpdf
   \usepackage[pdftex]{graphicx}

\else

\fi

\usepackage{amsmath}

\usepackage{array}

\usepackage{url}

\usepackage{textcomp}
\usepackage{booktabs}
\usepackage{threeparttable}

\begin{document}

\title{Characterization and Testing of a Micro-g Whispering Gallery Mode Optomechanical Accelerometer}

\author{Ying~Lia~Li,
        and~P.~F.~Barker%
\thanks{Y.~L.~Li and P.~F.~Barker are with the Department
of Physics \& Astronomy, University College London, London, WC1E 6BT, United Kingdom, e-mail: p.barker@ucl.ac.uk}%
\thanks{This work was supported in part by the EPSRC under EP/N509577/1, EP/H050434/1, \& EP/J500331/1.
}}
\markboth{This work has been submitted to the IEEE for possible publication.}%
{Shell \MakeLowercase{\textit{et al.}}: Bare Demo of IEEEtran.cls for IEEE Journals}

\maketitle

\begin{abstract}
Navigation, bio-tracking devices and gravity gradiometry are amongst the diverse range of applications requiring ultrasensitive measurements of acceleration. We describe an accelerometer that exploits the dispersive and dissipative coupling of the motion of an optical whispering gallery mode (WGM) resonator to a waveguide. A silica microsphere-cantilever is used as both the optical cavity and inertial test-mass. Deflections of the cantilever in response to acceleration alter the evanescent coupling between the microsphere and the waveguide, in turn causing a measurable frequency shift and broadening of the WGM resonance. The theory of this optomechanical response is outlined. By extracting  the dispersive and dissipative optomechanical rates from data we find good agreement between our model and sensor response. A noise density of 4.5\,\textmu g$\mathord{\cdot}$Hz$^{-\scriptscriptstyle\frac{1}{2}}$ with a bias instability of 31.8\,\textmu g (g\,=\,9.81\,m$\mathord{\cdot}$s$^{-2}$) is measured, limited by classical noise larger than the test-mass thermal motion. Closed-loop feedback is demonstrated to reduce the bias instability and long term drift. Currently this sensor outperforms both commercial accelerometers used for navigation and those in ballistocardiology for monitoring blood flowing into the heart. Further optimization would enable short-range gravitational force detection with operation beyond the lab for terrestrial or space gradiometry. 
\end{abstract}

\ifCLASSOPTIONpeerreview
\begin{center} \bfseries EDICS Category: 3-BBND \end{center}
\fi

\IEEEpeerreviewmaketitle

\section{Introduction}

\IEEEPARstart{S}{ignificant} progress in the field of optomechanics has stimulated a new generation of photonic motion detectors that surpass the sensitivity of current standards, with aims to approach fundamental limits set by quantum mechanics \cite{schliesser18}. These systems rely on the coupling between mechanical motion and light, which is enhanced when using an optical resonator such as a Fabry-Perot cavity \cite{taylor2014}, spherical cavities that support whispering gallery modes (WGMs) \cite{lia, pernice2009, painter, transduction, schliesser18, miao12, gavartin12, kippenbergg, haus}, and nanocavities in the form of photonic crystals \cite{barclay, krause}. Through optimization of the optical field geometry and the mechanical test-mass, sensitivities on the order of $10^{-18}$\,m$\mathord{\cdot}\mathrm{Hz}^{-\scriptscriptstyle\frac{1}{2}}$ \cite{ligo2016, schliesser18, rousseau06} can be reached, allowing laser cooling of macroscopic motion \cite{schliesser18, ulbricht}, force sensing \cite{gavartin12}, and recently the discovery of gravitational waves \cite{ligo2016}. 

Such high displacement sensitivities present opportunities to create optical inertial sensors such as accelerometers and gyroscopes. To date, only fiber-optic gyros and fiber Bragg grating (FBG) accelerometers have been commercialized. Typical FGBs are $10^{9}\times$ less sensitive than the current state-of-the-art set by the Laser Interferometer Gravitational-Wave Observatory (LIGO) \cite{natale}. LIGO's additional sensitivity is partly due to a dispersive optomechanical coupling where a movable end-mirror of the Fabry-Perot cavity shifts the resonance frequency. This provides higher signal-to-noise detection compared to schemes which monitor, for example, the efficiency of light coupled into a movable optical fiber test-mass \cite{pajak1998}. Dispersively coupled cavity optomechanical systems therefore show great promise for high-bandwidth sensing beyond 1\,kHz \cite{taylor2014, krause, miao12 } with additional benefits over their electrical counterparts such as immunity to electromagnetic interference. 

In this paper we use a spherical silica microcavity supporting optical WGMs that are attached to a silica cantilever to form a `microsphere-cantilever' test-mass. A waveguide is used to couple light to the WGM evanescently. Deflections of the microsphere-cantilever due to acceleration alter this evanescent coupling, perturbing the WGM resonance through a dispersive \emph{and} dissipative coupling that modulates the transmission through the waveguide. A benefit of dissipative coupling is high signal-to-noise sensing on-resonance. Unlike other WGM accelerometer schemes which detect an \emph{external} test-mass placed within the WGM's evanescent field \cite{miao12, gavartin12}, we use the WGM resonator as the test-mass itself to minimize extraneous mechanical degrees of freedom. This system, initially demonstrated by others some time ago \cite{haus}, lacked characterization of the accelerometer performance or specifications. To our knowledge, we report for the first time, a model governing the sensing mechanism as well as experimental data defining the sensing range, sensitivity, and stability via the Allan deviation. Such measurements are essential for applications such as navigation or gravity gradiometry that favor stability and linearity over short-term resolution. We achieve a noise density of 4.5\,\textmu g$\mathord{\cdot}$$\mathrm{Hz}^{-\scriptscriptstyle\frac{1}{2}}$ and bias instability of 31.8\,\textmu g; matching the performance of high grade accelerometers used for navigation \cite{ferrer, inertiallabs}. Closed-loop feedback further improves the bias instability and drift which are significant error sources for long term acceleration read-out. 

\section{Optomechanical Sensing of Acceleration}
The optical and mechanical properties of the WGM and microsphere-cantilever govern the accelerometer characteristics. We begin by describing the optical coupling to the WGM before defining the mechanical response of the microsphere-cantilever test-mass. 
\subsection{Evanescent Coupling of Whispering Gallery Modes}
\begin{figure}[!t]
\centering
\includegraphics[width=8.8cm]{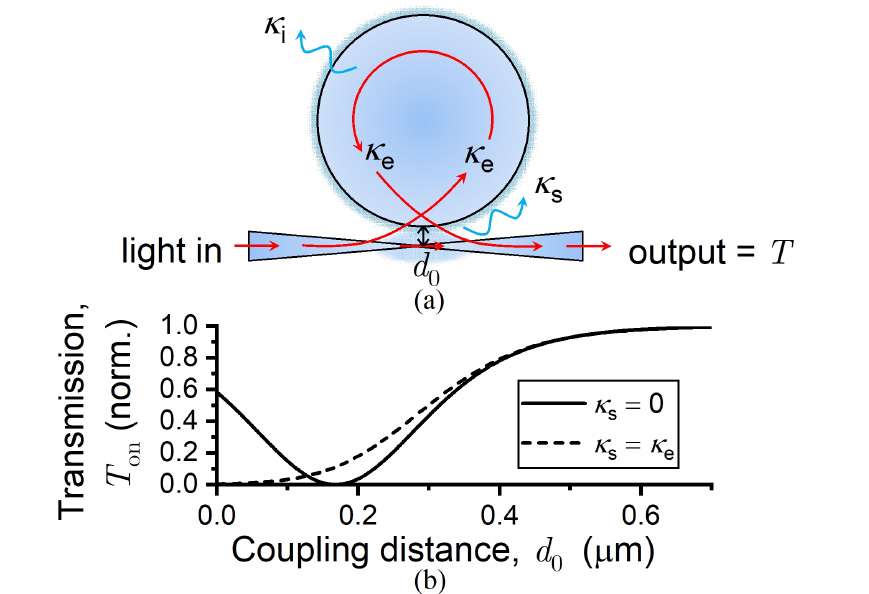}
\caption{(a) Evanescent coupling of light from a tapered waveguide into a WGM at a coupling distance $d_{\mathrm{0}}$. The transmission, $T$, depends on three coupling rates; the intrinsic, $\kappa_{\mathrm{i}}$, the extrinsic, $\kappa_{\mathrm{e}}$, and a scattering rate, $\kappa_{\mathrm{s}}$. (b) $\kappa_{\mathrm{e}}$ and $\kappa_{\mathrm{s}}$ vary exponentially with $d_{0}$ which affects the on-resonance transmission, $T_{\mathrm{on}}$ (shown as a simulated plot). When $\kappa_{\mathrm{s}}\geq\kappa_{\mathrm{e}}$, $T_{\mathrm{on}}$ no longer shows a turning point at the critical coupling distance $d_{\rm{c}}$ defined by $T_{\rm{on}}=0$ when $\kappa_{\rm{s}}=0$.  }
\label{coupling}
\end{figure}

The evanescent field of a tapered optical fiber with a waist smaller than the input light wavelength is overlapped with the evanescent field of a WGM cavity, enabling light transfer \cite{vahala}. The transmission of the taper waveguide, $T$, is related to the power coupled to the WGM, $P_{\mathrm{c}}$, by $T=1-P_{\mathrm{c}}$ and is governed by three coupling rates; the extrinsic coupling $\kappa_{\mathrm{e}}$ that defines light transfer from taper to WGM, the intrinsic coupling $\kappa_{\mathrm{i}}$, and a scattering component $\kappa_{\mathrm{s}}$, shown in Fig.~\ref{coupling}\,(a). The material and surface quality of the resonator limits $\kappa_{\mathrm{i}}$, whereas $\kappa_{\mathrm{s}}$ accounts for optical losses that do not couple back into the waveguide \cite{transduction}.

Using coupled-mode theory \cite{haus1}, the WGM intracavity electromagnetic field, $a$, is expressed by:
\begin{equation}
\frac{\mathrm{d}a}{\mathrm{d}t}=-\Big(\frac{\kappa_{\mathrm{i}}}{2}+\frac{\kappa_{\mathrm{e}}}{2}+\frac{\kappa_{\mathrm{s}}}{2}+i\Delta\Big)a+\sqrt{\kappa_{\mathrm{e}}}a_{\mathrm{in}},
\end{equation}
where $a_{\mathrm{in}}$ is defined by the laser input power, $P_{\mathrm{in}}=a^{2}_{\mathrm{in}}\hbar\omega$, and where $\Delta=\omega-\omega_{0}$ is the detuning of the laser from the WGM resonance frequency $\omega_{0}$. The optical quality factor is defined by $Q_{\mathrm{opt}}=\frac{\omega_{0}}{\kappa}$ where $\kappa=\kappa_{\mathrm{i}}+\kappa_{\mathrm{e}}+\kappa_{\mathrm{s}}$. The output field of the waveguide follows the input-output relationship $a_{\mathrm{out}}=-a_{\mathrm{in}}+\sqrt{\kappa_{\mathrm{e}}}a$. Since the cavity photon lifetime is many orders of magnitude shorter than the acceleration timescales to be considered in this paper, the steady-state condition is applied such that $\frac{\mathrm{d}a}{\mathrm{d}t}=0$. The normalized transmission can therefore be expressed as $T=|\frac{a_{\mathrm{out}}}{a_{\mathrm{in}}}|^{2}$:
\begin{equation}
T=\left|{1-\frac{\kappa_{\mathrm{e}}}{\frac{\kappa_{\mathrm{i}}}{2}+\frac{\kappa_{\mathrm{e}}}{2}+\frac{\kappa_{\mathrm{s}}}{2}+i\Delta}}\right|^{2}.
\label{coup}
\end{equation}
The exponential decay length of the evanescent field of the waveguide and sphere, $\alpha_{\mathrm{w}}$ \& $\alpha_{\mathrm{r}}$ respectively, cause $\kappa_{\mathrm{e}}$ and $\kappa_{\mathrm{s}}$ to vary with the coupling distance, marked by $d_{\mathrm{0}}$ in Fig.~\ref{coupling}\,(a). The coupling varies as $e^{-\eta d_{\mathrm{0}}}\approx e^{-\frac{d_{\mathrm{0}}}{\alpha_{\mathrm{w}}+\alpha_{\mathrm{r}}}}$ \cite{haus1}. The transmission therefore changes with $d_{\mathrm{0}}$, as shown in Fig.~\ref{coupling}\,(b) for the on-resonance transmission, $T_{\mathrm{on}}$, when $\Delta=0$. The critical coupling position, $d_{\mathrm{c}}$, occurs when $\kappa_{\mathrm{i}}=\kappa_{\mathrm{e}}$, as indicated when $T_{\mathrm{on}}=0$ with $\kappa_{\rm{s}}=0$. When $\kappa_{\rm{s}}=\kappa_{\rm{e}}$, the turning point of $T_{\rm{on}}$ for $d_{\rm{0}}<d_{\rm{c}}$ is no longer present.

\subsection{Measurement of Displacement \& Acceleration}
\begin{figure}[!t]
\centering
\includegraphics[width=8.8cm]{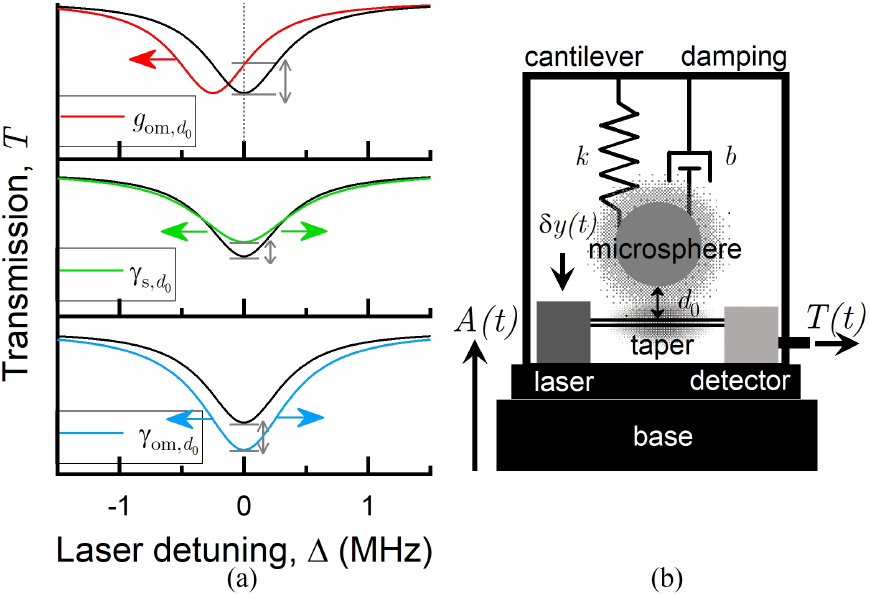}
\caption{(a) The WGM (black) is modified when $d_{\mathrm{0}}$ changes, governed by three optomechanical rates; the dispersive rate g$_{\mathrm{om}}$, the scattering rate $\mathrm{\gamma_{s}}$, and the dissipative rate $\mathrm{\gamma_{om}}$. Arrows indicate the resultant simulated shift of the resonant frequency (top panel) and linewidth broadening (middle \& bottom panels). (b) The WGM accelerometer schematic. The test-mass is the microsphere-cantilever with spring constant $k$ that deflects by $\delta y(t)$ when the base is accelerated by $A(t)$. Damping, $b$, is provided by the surrounding air that fills the chamber.}
\label{mechanism}
\end{figure}

Changes to $d_{\mathrm{0}}$ create three effects on the WGM lineshape, depicted in Fig.~\ref{mechanism}\,(a); a dispersive shift in $\omega_{\rm{0}}$ and two dissipative losses that broaden the linewidth. All three modulate the intracavity power, $P_{\mathrm{c}}$. An optomechanical rate describes the magnitude of each effect.
The broadening of the WGM linewidth due to a change in $d_{\rm{0}}$ is given by $\gamma_{\mathrm{om}}=\frac{\mathrm{d}\kappa_{\rm{e}}}{\mathrm{d}d_{\rm{0}}}$, where $\kappa_{\mathrm{e}}=\kappa_{\mathrm{e,0}}e^{-\eta d_{\mathrm{0}}}$, and $\kappa_{\mathrm{e,0}}$ is the `zero-gap' coupling rate \cite{transduction, pernice2009, barclay}. By considering small displacements around a fixed $d_{\mathrm{0}}$, one can linearize $\gamma_{\mathrm{om}}$ as $\gamma_{\mathrm{om}}(d_{\mathrm{0}})=\frac{\mathrm{d}\kappa_{\mathrm{e}}(d_{\mathrm{0}})}{\mathrm{d}d_{\mathrm{0}}}$. Further broadening is provided by $\gamma_{\mathrm{s}}(d_{\mathrm{0}})=\frac{\mathrm{d}\kappa_{\mathrm{s}}(d_{\mathrm{0}})}{\mathrm{d}d_{\mathrm{0}}}$, derived from $\kappa_{\mathrm{s}}(d_{\mathrm{0}})=\kappa_{\mathrm{s,0}}e^{-\eta d_{\mathrm{0}}}$. The dispersive optomechanical rate describes the red-shift of the WGM resonance due to the influence of the tapered fiber on the local effective refractive index near the surface of the WGM resonator \cite{kippenbergg}. This is given by $g_{\mathrm{om}}(d_{\mathrm{0}})=\frac{\mathrm{d}\Delta(d_{\mathrm{0}})}{\mathrm{d}d_{\mathrm{0}}}$ where $\Delta(d_{\mathrm{0}})=\Delta^{*}-\Delta_{\mathrm{0}}e^{-\eta d_{\mathrm{0}}}$ and $\Delta^{*}$ is the unperturbed detuning.

We now describe the operation of this system as an accelerometer. An acceleration, $A(t)$, in the vertical direction, is applied to the base of the setup causing the cantilever to deflect by a distance $\delta y(t)$ away from $d_{\rm{0}}$, as illustrated in Fig.~\ref{mechanism}\,(b). Hooke's law equates $\delta y(t)$ with the spring constant $k$ and the effective mass of the microsphere-cantilever, $m_{\mathrm{eff}}$:
\begin{equation}
\label{mech}
A(t)=-\frac{k}{m_{\mathrm{eff}}}\delta y(t)=-\Omega_{\mathrm{m}}^{2}\delta y(t),
\end{equation}
where $\Omega_{\mathrm{m}}$ is the fundamental mechanical resonance frequency of the microsphere-cantilever identified in the power spectral density (PSD=$S(f)$) of $T$ as previously outlined in \cite{lia}. Tailoring the geometry of the microsphere-cantilever can tune $\Omega_{\mathrm{m}}$ by considering Euler-Bernoulli beam theory for a cantilever with an optical cavity end-mass $m_{\mathrm{s}}$:
\begin{equation}
\Omega_{\mathrm{m}}=\sqrt{\frac{3EI}{(0.24\rho SL+m_{\mathrm{s}})L^{3}}}=\sqrt{\frac{k}{m_{\mathrm{eff}}}},
\label{mecheq}
\end{equation} 
where $k=\sqrt{\frac{EI}{L^{3}}}$, $E$ is Young's modulus of silica, $I$ is the second moment of area for the uniform cross-section of the cantilever, $\rho$ is the density of silica, $S$ is the cross-sectional area of the cantilever, and $L$ is the cantilever length.  

Detection of $\delta y(t)$ due to the change in $T(t)$ can be modeled by considering dispersive and dissipative transduction \cite{barclay}, where:
\begin{equation}
\label{full}
\mathrm{d}T(t)=\left|g_{\mathrm{om}}\frac{\partial T}{\partial\Delta}+\gamma_{\mathrm{om}}\frac{\partial T}{\partial\kappa_{\mathrm{e}}}+\gamma_{\mathrm{s}}\frac{\partial T}{\partial\kappa_{\mathrm{s}}}\right|\mathrm{d}y(t).
\end{equation}
For the WGM system studied here;
\begin{equation}
\label{eqe}
\frac{\partial T}{\partial\kappa_{\mathrm{e}}} = -\frac{4(\kappa_{\mathrm{i}}+\kappa_{\mathrm{s}})(4\Delta^{2}-\kappa_{\mathrm{e}}^{2}+(\kappa_{\mathrm{i}}+\kappa_{\mathrm{s}})^{2})}{(4\Delta^{2}+\kappa^{2})^{2}},
\end{equation}
\begin{equation}
\label{eqd}
\frac{\partial T}{\partial\Delta} = \frac{32\Delta\kappa_{\mathrm{e}}(\kappa_{\mathrm{i}}+\kappa_{\mathrm{s}})}{(4\Delta^{2}+\kappa^{2})^{2}}, 			\mathrm{and}
\end{equation}
\begin{equation}
\label{eqs}
\frac{\partial T}{\partial\kappa_{\mathrm{s}}} = -\frac{4\kappa_{\mathrm{e}} (4\Delta^{2}+\kappa_{\mathrm{e}}^{2}-(\kappa_{\mathrm{i}}+\kappa_{\mathrm{s}})^{2})}{(4\Delta^{2}+\kappa^{2})^{2}}.
\end{equation}
The interplay between these derivatives and the ratio between $g_{\mathrm{om}}, \gamma_{\rm{om}}$, and $\gamma_{\rm{s}}$ defines the scale-factor at each $d_{\mathrm{0}}$, i.e. the change in $T$ per metre. In particular, (\ref{eqe}) \& (\ref{eqs}) are unipolar and symmetrical whereas (\ref{eqd}) is bipolar and asymmetrical. We note that an increase in scale-factor does not necessarily yield an improvement in sensitivity as noise can also be amplified. 

The \emph{measured} linear scale-factor recorded by a photodetector monitoring $T$ (in units of Volts/g where g\,=\,9.81\,m$\mathord{\cdot}$s$^{-2}$) may be much smaller than the sensing range governed by Hooke's law due to the optomechanical detection. Understanding the non-linearity of the scale-factor as well as the sensor stability, previously unreported, forms the basis of the proceeding sections.

\section{Characterization \& testing} 
First we describe the experimental set-up and determine the optical properties of the system. Next we characterize the WGM accelerometer in terms of the linear sensing range, scale-factor, sensitivity and stability. Lastly, a simple closed-loop feedback scheme is shown, providing a reduction in noise and drift over long timescales as a proof-of-concept for application driven requirements. 
\subsection{Experimental Details}
\label{experiment}
\begin{figure}[!t]
\centering
\includegraphics[width=8.8cm]{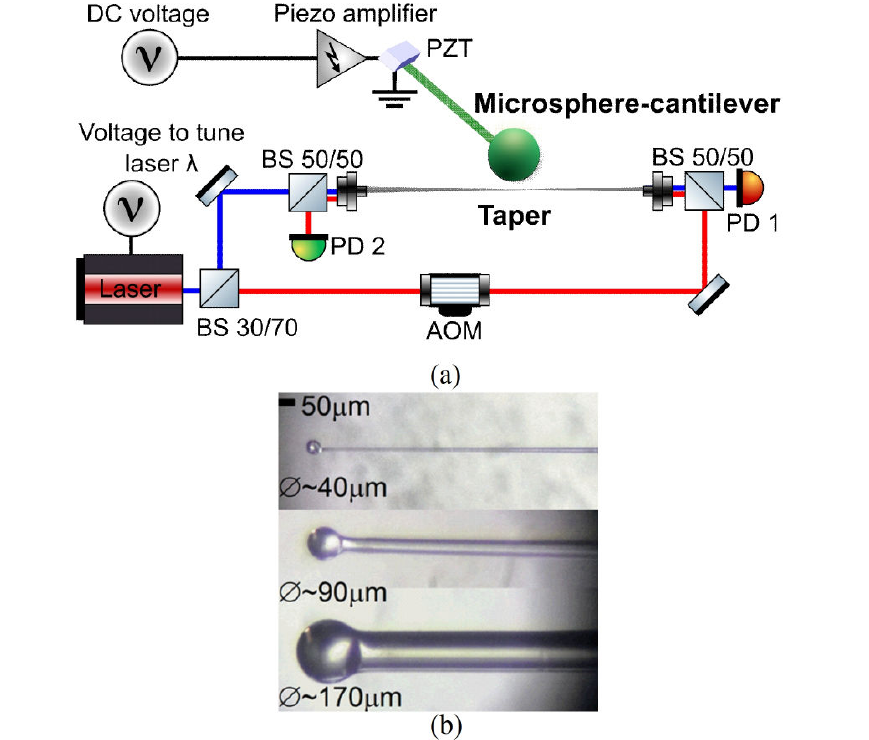}
\caption{(a) Experimental setup where the microsphere-cantilever is attached to a piezostack (PZT) to tune the coupling distance $d_{\mathrm{0}}$. A WGM is excited on-resonance by launching a tunable laser beam (blue beam) into a tapered fiber whose transmission $T$ is measured by photodetector PD\,1. A beamsplitter (BS) allows for a counterpropagating beam (red beam) to be launched and detected by PD\,2 with a detuning of $\Delta$ controlled by two acousto-optic modulators (AOM). (b) Various diameter ($\oslash$) microsphere-cantilevers made by melting standard or pre-tapered optical fiber.}
\label{setup}
\end{figure}

The microsphere-cantilever is fabricated by melting the tip of an optical fiber using a CO$_{2}$ laser, allowing surface tension to form a sphere \cite{lia}. Evanescent waveguides known as tapers are created by pulling a length of fiber whilst heating with a butane flame to achieve a minimum waist of approximately 1\,\textmu m \cite{lia}. A homebuilt tunable diode pumped Nd:YVO$_{4}$ 1064\,nm laser is used to couple light into the waveguide. A piezostack (PZT) at the clamped end of the microsphere-cantilever controls $d_{\mathrm{0}}$. In Fig.~\ref{setup}\,(a) is shown the experimental setup, with a microscope image of typical sized microsphere-cantilevers displayed in (b). The microsphere-cantilever and taper are housed within a sealed chamber, at atmospheric pressure. 

Obtaining a high signal-to-noise measurement of $\mathrm{d}T$ requires the laser to sit on the detuned slope of the WGM resonance for dispersively coupled systems. Previously, this was achieved using a Pound-Drever Hall lock \cite{lia}. However, this becomes unstable in the presence of driven vibrations and a passive lock is employed instead on WGM resonances with $>50\,$MHz linewidths \cite{carmon}. This locking beam, shown as the blue beam in Fig.~\ref{setup}\,(a), is wavelength tuned on-resonance ($\Delta=0$) by applying a voltage to the laser until the transmission, measured by photodetector PD\,1, reaches a minimum. A counterpropagating WGM is excited by launching light through the other end of the taper to create a transduction beam (red beam). This is detuned with respect to the locking beam by $\Delta=140\,$MHz using acousto-optic modulators (AOM) whose transmission is monitored by PD\,2. After adjustment of $d_{\mathrm{0}}$ via the PZT, the locking beam is re-scanned to ensure $T_{\rm{on}}$ remains at a minimum. 

\begin{figure}[!t]
\centering
\includegraphics[width=8.8cm]{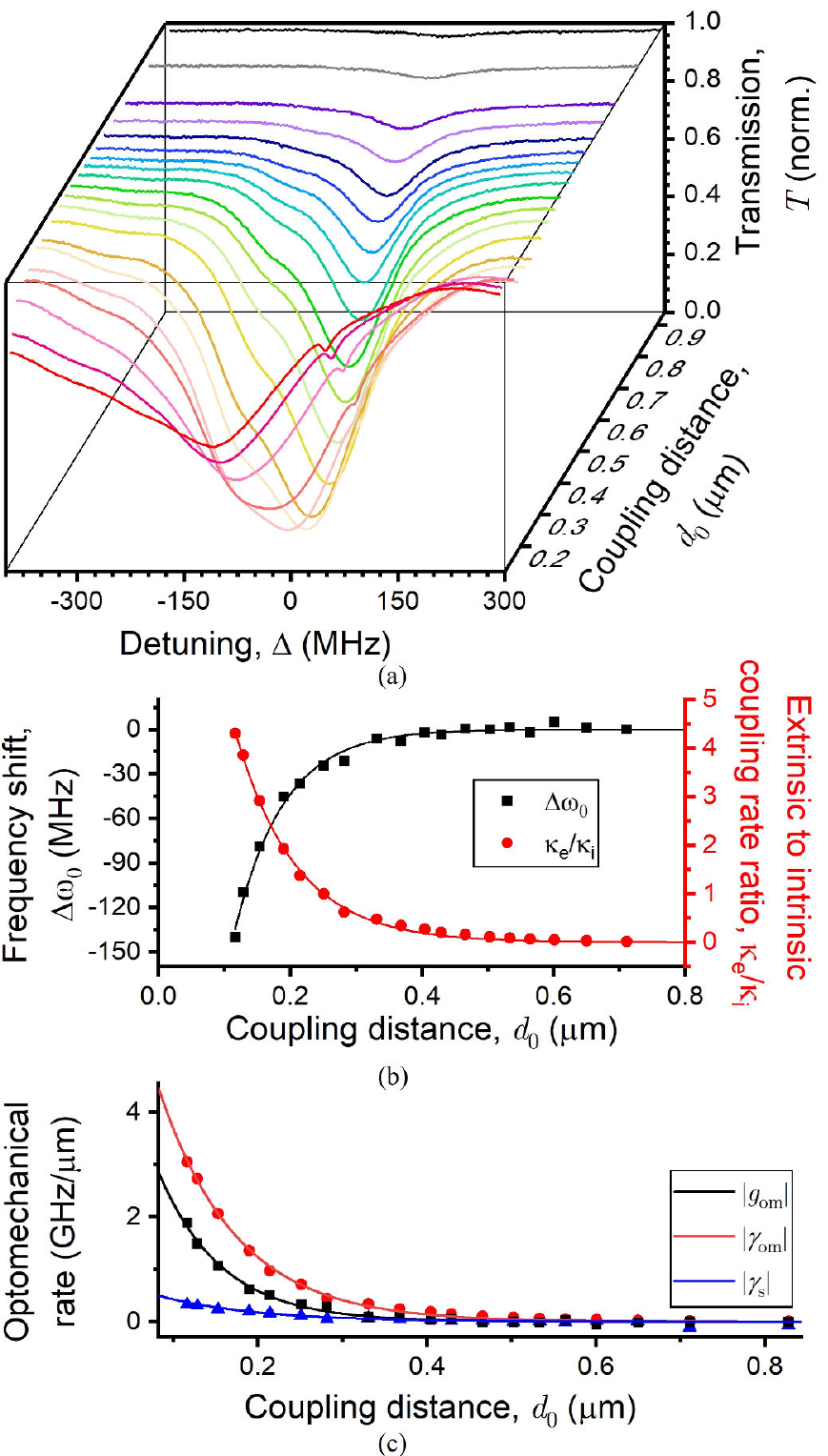}
\caption{(a) $T$ as a function of $d_{\rm{0}}$ as the laser is scanned across the WGM where $\Delta=0$ indicates the unshifted resonance frequency $\omega_{0}$. (b) Plotting $\kappa_{\rm{e}}$ and $\Delta$ extracted from (a) provides $g_{\mathrm{om}}(d_{\mathrm{0}})=\frac{\mathrm{d}\Delta(d_{\mathrm{0}})}{\mathrm{d}d_{\mathrm{0}}}$ and $\gamma_{\mathrm{om}}(d_{\mathrm{0}})=\frac{\mathrm{d}\kappa_{\mathrm{e}}(d_{\mathrm{0}})}{\mathrm{d}d_{\mathrm{0}}}$ respectively. In (c) is shown the magnitude of the optomechanical rates as a function of $d_{\rm{0}}$ where $\gamma_{\mathrm{s}}(d_{\mathrm{0}})=\frac{\mathrm{d}\kappa_{\mathrm{s}}(d_{\mathrm{0}})}{\mathrm{d}d_{\mathrm{0}}}=\frac{\rm{d}}{\mathrm{d}d_{\rm{0}}}(\delta\omega_{\rm{0}}-\kappa_{\rm{e}}-\kappa_{\rm{i}})$, and $\delta\omega_{\rm{0}}$ is the FWHM linewidth. Lines indicate fits whereas points are experimental data.}
\label{rates}
\end{figure}

The microsphere-cantilever studied primarily in this paper has a microsphere diameter of 120\,\textmu m, cantilever diameter of 80\,\textmu m and cantilever length of 22\,mm. To extract $g_{\mathrm{om}}(d_{\mathrm{0}})=\frac{\mathrm{d}\Delta(d_{\mathrm{0}})}{\mathrm{d}d_{\mathrm{0}}}$, $\gamma_{\mathrm{om}}(d_{\mathrm{0}})=\frac{\mathrm{d}\kappa_{\mathrm{e}}(d_{\mathrm{0}})}{\mathrm{d}d_{\mathrm{0}}}$ and $\gamma_{\mathrm{s}}(d_{\mathrm{0}})=\frac{\mathrm{d}\kappa_{\mathrm{s}}(d_{\mathrm{0}})}{\mathrm{d}d_{\mathrm{0}}}$, the taper transmission is first recorded as a function of $d_{\rm{0}}$ whilst the laser is scanned across the WGM, as shown in Fig.~\ref{rates}\,(a). We note that scattering plays only a small role due to the presence of a turnaround where $T_{\rm{on}}$ increases with decreasing $d_{\rm{0}}$ (turnaround ceases if $\kappa_{\rm{s}}=\kappa_{\rm{e}}$, Fig.~\ref{coupling}\,(b)).
Due to the presence of $g_{\rm{om}}$, $\gamma_{\rm{om}}$ and $\gamma_{\rm{s}}$, a change in $d_{\rm{0}}$ decreases the detuning $\Delta$ by red-shifting $\omega_{0}$ and broadens the linewidth, $\delta\omega_{\rm{0}}$, from the intrinsic FWHM linewidth of 65\,MHz. By plotting $\Delta$ and $\kappa_{\rm{e}}$ as a function of $d_{\rm{0}}$ in Fig.~\ref{rates}\,(b), one can extract $g_{\rm{om}}$ and $\gamma_{\rm{om}}$ respectively. Scattering creates an additional broadening unaccounted by $\kappa_{\rm{e}}$ therefore $\gamma_{\rm{s}}=\frac{\rm{d}}{\mathrm{d}d_{\rm{0}}}(\delta\omega_{\rm{0}}-\kappa_{\rm{e}}-\kappa_{\rm{i}})$. 

The optical coupling rates are $\kappa_{\rm{e}}=1.013\times 10^{9}e^{-10.9\times 10^{-6}d_{\rm{0}}}$\,Hz, $\kappa_{\rm{s}}=0.117\times 10^{9}e^{-8.8\times 10^{-6}d_{\rm{0}}}$\,Hz, and the redshift is $\Delta=0.649\times 10^{9}e^{-13.5\times 10^{-6}d_{\rm{0}}}$\,Hz. The magnitude of the three optomechanical rates are shown in Fig.~\ref{rates}\,(c). The dispersive, dissipative and scattering optomechanical rates are $g_{\rm{om}}=8.70e^{-13.4d_{\rm{0}}}$\,GHz$\mathord{\cdot}$\textmu m$^{-1}$, $\gamma_{\rm{om}}=11.04e^{-10.9d_{\rm{0}}}$\,GHz$\mathord{\cdot}$\textmu m$^{-1}$ and $\gamma_{\rm{s}}=1.03e^{-8.8d_{\rm{0}}}$\,GHz$\mathord{\cdot}$\textmu m$^{-1}$ respectively, noting that $d_{\rm{0}}$ is measured in micrometres. The different decay length $\eta$ for all three optomechanical rates will be related to the unknown origin of the scattering rate and extraneous red-shifting of the WGM due to thermal effects \cite{transduction}. 

\subsection{Sensing Range \& Scale-factor}
\begin{figure*}[!t]
\centering
\includegraphics[width=18.1cm]{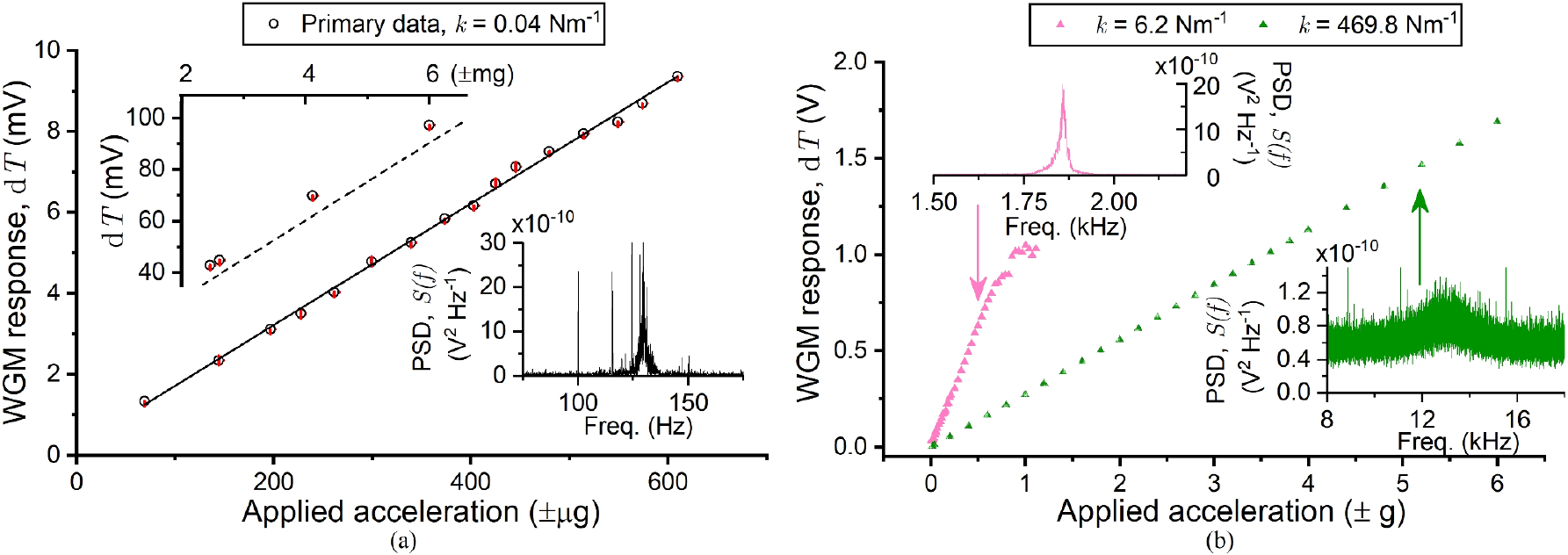}
\caption{(a) Linear sensing range of the WGM accelerometer with a scale-factor of 15.00\,V/g$\pm$\,0.22\,V/g for $A<\pm 700$\,\textmu g, found by applying a linear fit to the WGM tranduction beam response where the photodetector is amplified and AC coupled. Data at higher $A$, top inset, results in a large deviation from the fit (dashed line). The PSD of the accelerometer output at rest (photodetector DC coupled with no additional gain), bottom inset, provides $\Omega_{\rm{m}}$ and $Q_{\rm{m}}$. Using equation (\ref{mecheq}), $k=0.04$\,N$\mathord{\cdot}$m$^{-1}$. (b) A larger sensing range is achieved by tailoring the microsphere-cantilever dimensions (see Table~\ref{mechparam}) producing $k=6.2$\,N$\mathord{\cdot}$m$^{-1}$ (pink data, linear range $\pm 0.7\,$g) and $k=469.8$\,N$\mathord{\cdot}$m$^{-1}$ (green data, linear range $\pm 6$\,g but limited by shaker output).  }
\label{range}
\end{figure*}

A sinusoidal shake test is performed to provide an applied acceleration of $A(t)=D\Omega^{2}\sin(\Omega t)$ where $D$ is the shake amplitude and $\Omega$ is shake frequency. A commercial sensor made by Analog Devices (ADXL327, range $\pm$2\,g) is used for calibration and a sine fit function is used to extract the WGM response with the photodetector AC coupled with additional gain. Shown in the main plot of Fig.~\ref{range}\,(a) is the WGM response to $A(t)<\pm 700\,$\textmu g at $\Omega=2\pi\times 83$\,Hz, with a 10,000 point Savitzky-Golay filter used to reduce white noise. The gradient of the fit provides a linear scale-factor of 15.00\,V/g$\pm$\,0.22\,V/g for $A<\pm 700$\,\textmu g. Further increasing $A$, shown in the top inset of Fig.~\ref{range}\,(a), results in an increasing deviation from the linear fit until the microsphere-cantilever touches the taper around $A>\pm 8$\,mg. By fitting the peak in the PSD (bottom inset in Fig.~\ref{range}\,(a)), the mechanical resonance frequency is $\Omega_{\rm{m}}=2\pi\times 133$\,Hz which is used in equation (\ref{mecheq}) to find $k=0.04$\,N$\mathord{\cdot}$m$^{-1}$. The mechanical linewidth is determined to be $\delta\omega_{\rm{0,m}}=2\pi\times 3\,$Hz, leading to a mechanical quality factor of $Q_{\rm{m}}=\frac{\Omega_{\rm{m}}}{\delta\omega_{\rm{0,m}}}=44.3$ \cite{lia}.

In Fig.~\ref{range}\,(b) is shown the sensing range of two alternative microsphere-cantilever test-masses with larger $k$ from 6.2\,N$\mathord{\cdot}$m$^{-1}$ (pink data) to 469.8\,N$\mathord{\cdot}$m$^{-1}$ (green data). Simply adjusting the mechanical dimensions can therefore create accelerometers suitable for higher accelerations exceeding $\pm 6$\,g. Table~\ref{mechparam} lists the different dimensions and the mechanical properties obtained from the PSD for each system (Fig.~\ref{range}\,(b) inset graphs).

\begin{table*}[!t]
\centering
\begin{threeparttable}[b]
\renewcommand{\arraystretch}{1}
\caption{Dimensions \& mechanical properties of 3 microsphere-cantilevers test-masses.}
\label{mechparam}
\centering
\begin{tabular}{l c c c c c c}
\toprule
Data set in Fig.~\ref{range} & Microsphere $\oslash$ (\textmu m) & Cantilever cross--section (\textmu m)  & Cantilever $L$ (mm) & $\frac{\Omega_{\rm{m}}}{2\pi}$ (Hz)  & $Q_{\rm{m}}$ & $k$ (N$\mathord{\cdot}$m$^{-1}$)   \\
\midrule
Black (primary) & 120 & ($\oslash$) 80 & 22 & 133 & 44.3 & 0.04 \\
Pink  & 152 & ($\oslash$) 120  & 7 & 1860 & 84.5 & 6.2\\
Green & 350 & ($b\times h$) 250\,$\times$106 & 2.2 & 13160 & 5.3 & 469.8  \\
\bottomrule
\end{tabular}
\begin{tablenotes}
\footnotesize
\item $\oslash$ indicates diameter whereas $b$ and $h$ are the base and height of a rectangular-core optical fiber. $\Omega_{\rm{m}}$ is the fundamental mechanical resonance frequency, $Q_{\rm{m}}$ is the mechanical quality factor, and $k$ the spring constant. 
\end{tablenotes}
\end{threeparttable}
\end{table*}

\begin{figure}[!t]
\centering
\includegraphics[width=8.8cm]{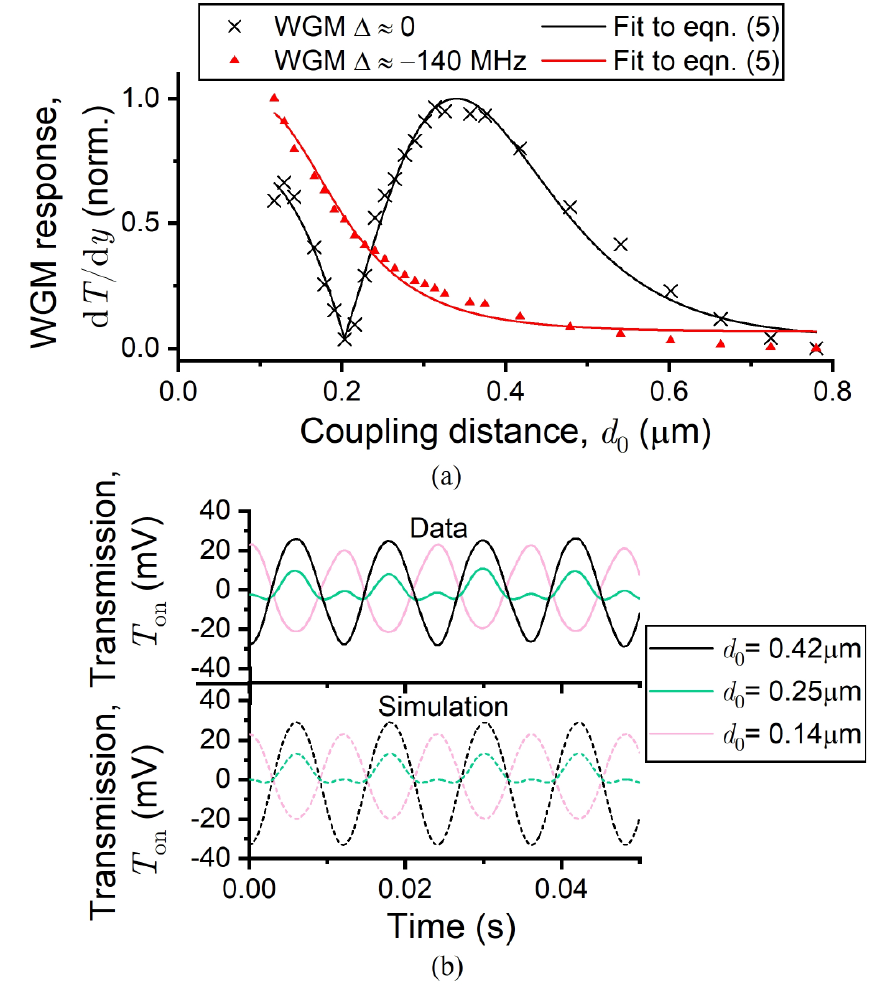}
\caption{(a) Normalized scale-factor for the locking beam (black crosses) and transduction beam (red triangles) as a function of $d_{\mathrm{0}}$. Both data are fitted to equation (\ref{full}). (b) Top panel: Time trace of $T_{\rm{on}}$ (locking beam data) during the shake test at varying $d_{\rm{0}}$. Bottom panel: Simulated $T_{\rm{on}}$ at the corresponding $d_{\rm{0}}$ to the top panel. }
\label{scalefactor}
\end{figure}

The deviation from the linear response in the inset graph of Fig.~\ref{range}\,(a) is due to the exponential dependence of the optomechanical rates with $d_{\mathrm{0}}$. The same effect is seen in Fig.~\ref{range}\,(b) on the pink data set for a microsphere-cantilever with $k=6.2$$\mathord{\cdot}$N\,m$^{-1}$. This non-linearity can be compared with the model outlined in equation (\ref{full}) by measuring the scale-factor whilst adjusting $d_{\mathrm{0}}$. This is conducted for the primary system and shown in Fig.~\ref{scalefactor}\,(a) where the normalized, relative scale-factor for the detuned transduction beam ($\Delta=140\,$MHz) and the passively locked on-resonance beam ($\Delta=0$\,MHz) is plotted as a function of $d_{\mathrm{0}}$.

The fits in Fig.~\ref{scalefactor}\,(a) use equation (\ref{full}) with known values for the optical coupling rates ($\kappa_{\mathrm{i}}$, $\kappa_{\rm{e}}$, $\kappa_{\rm{s}}$) and the optomechanical rates ($g_{\rm{om}}$, $\gamma_{\rm{om}}$, $\gamma_{\rm{s}}$) as measured earlier in Fig.~\ref{rates} and detailed in section~\ref{experiment}. The remaining free parameter is the detuning which is found to be $1\,$MHz$\pm 1\,$MHz for the locking beam and $-146$\,MHz$\pm 8$\,MHz for the transduction beam, in good agreement with the experimental settings.  
Deviation of the fit to data points when $d_{\mathrm{0}}>500\,$nm is due to calibration error of $d_{\mathrm{0}}$ from the point of contact between microsphere-cantilever and taper, i.e. $d_{\mathrm{0}}=0$. The relative voltage applied to the PZT at each step is used to calculate $\Delta d_{\mathrm{0}}$ but assumes PZT linearity and no drift. The accuracy should be improved by using a strain gauge PZT in closed-loop mode to provide an absolute measurement of $d_{\mathrm{0}}$. Increased stability of laser locking to the WGM will also enhance measurement of the scale-factor change with $d_{\rm{0}}$. It should be noted that despite the scale-factor dependence on $d_{\mathrm{0}}$, linearization of $\frac{\mathrm{d}T}{\mathrm{d}y}$ about a null-position $d_{\mathrm{0}}$ remains valid, as confirmed in Fig.~\ref{range}. 

Although there is an exponential improvement to the optomechanical rates with decreasing $d_{\rm{0}}$, the turnaround point of $T_{\rm{on}}$ at $d_{\rm{c}}\approx 0.25\,$\textmu m heavily distorts vibration signals such as the sinusoidal acceleration profile in Fig.~\ref{scalefactor}\,(b). The top panel is the experimental locking beam signal whereas the bottom panel shows a simulation of $T_{\rm{on}}$ using equation (\ref{coup}) with $\Delta=0$, $\kappa_{\rm{i}}=65\,$MHz, $\kappa_{\rm{e}}=1.013\times10^{9}e^{-10.9\times 10^{6}(d_{\rm{0}}-d\cos{(\Omega t)})}$\,Hz, and $\kappa_{\rm{s}}=0.117\times10^{9}e^{-8.8\times 10^{6}(d_{\rm{0}}-d\cos{(\Omega t)})}$\,Hz, where $d$ is the modeled deflection of the cantilever. Close to $d_{\rm{c}}$ (green data), the sinusoidal response becomes deformed from the expected response (black data). A similar effect occurs when $T$ is detuned but at a different $d_{\rm{0}}$. In general, operating the sensor at $d_{\rm{0}}\ll d_{\rm{c}}$ is not recommended as it creates an inverted response where $T$ increases with decreasing $d_{\rm{0}}$ (Fig.~\ref{scalefactor}\,(b), pink data), whilst increasing the likelihood of the WGM cavity touching the taper waveguide.

\subsection{Sensitivity \& Noise}
The fundamental limit of sensitivity for all accelerometers utilizing a test-mass is the thermal motion of this mass, known as the thermal noise equivalent acceleration $a_{\rm{th}}$:
\begin{equation}
a_{\rm{th}}=\sqrt{\frac{4k_{\rm{B}}T_{\rm{0}}\Omega_{\rm{m}}^{3}}{kQ_{\rm{m}}}},
	\label{neaequation}
	\end{equation} 
where $k_{\rm{B}}$ is Boltzmann's constant, and $T_{\rm{0}}$ the mode temperature of the fundamental mechanical resonance. The \emph{total} noise equivalent acceleration $a_{\rm{nea}}$ is defined as $a_{\rm{nea}}=\sqrt{a_{\rm{th}}^{2}+a^{2}_{\rm{det}}+a^{2}_{\rm{add}}}$ where $a_{\rm{th}}$ arises from the microsphere-cantilever thermal motion, $a_{\rm{det}}$ is related to readout noise, and $a_{\rm{add}}$ is a lumped parameter of other noise sources such those from the laser and electronics. Since $a_{\rm{det}}$ and $a_{\rm{add}}$ can be actively minimized, $a_{\rm{th}}$ represents the best obtainable limit without exploiting a quantum test-mass. For the parameters extracted in Table~\ref{mechparam}, the WGM sensor limit is $a_{\rm{th}}=0.24$\,\textmu g$\mathord{\cdot}$Hz$^{-\scriptscriptstyle\frac{1}{2}}$.

The Allan deviation, $\sigma$, allows extraction of $a_{\rm{nea}}$ by characterizing noise and stability as a function of sampling time, $\tau$. It is a time-domain metric computed from a long sample of data recording the accelerometer output with the sensor at rest and is related to the PSD $S(f)$ by \cite{ferrer}:
\begin{equation}
\sigma(\tau)=2\bigg[\int_{0}^{\infty}S(f)\frac{\sin^{4}(\pi f\tau)}{(\pi f\tau)^{2}}\,\mathrm{d}f\bigg]^{1/2}.
	\label{allanpsd}
	\end{equation} 

\begin{figure}[h!]
\centering
\includegraphics[width=8.8cm]{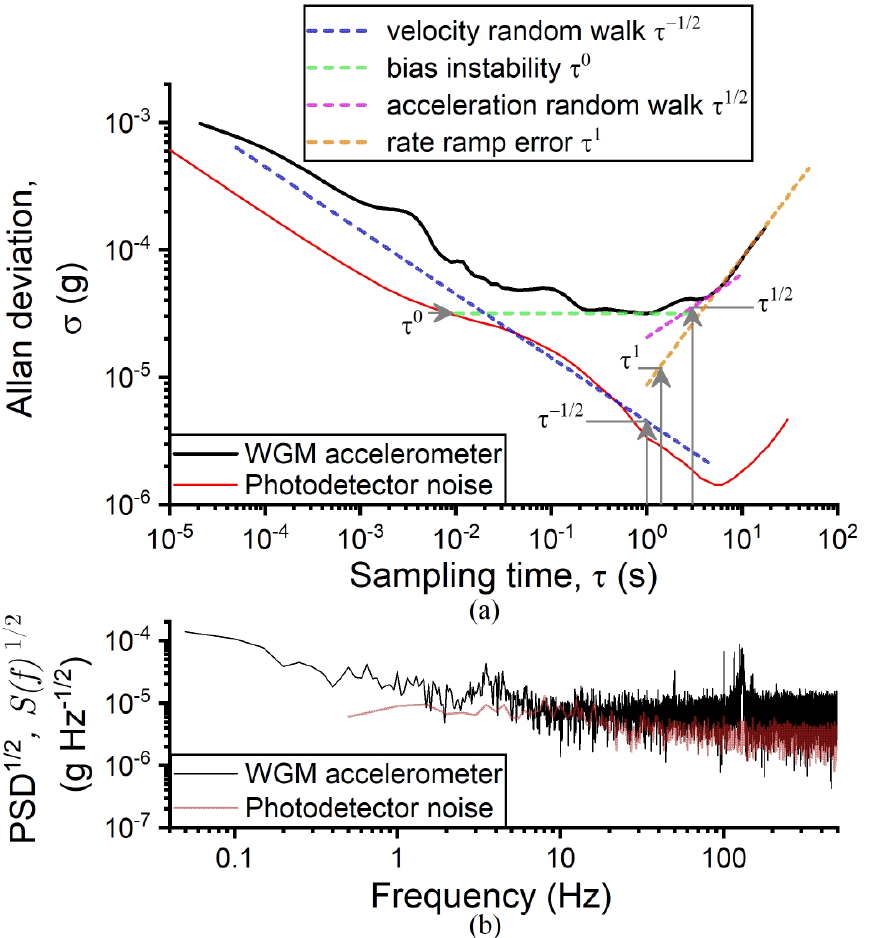}
\caption{(a) The Allan deviation, $\sigma$, as a function of sampling time, $\tau$, for the WGM accelerometer (black) and photodetector with no light (red). The noise profiles are identified by their differing gradients on the Allan deviation. The magnitude, $|\sigma|$, of the VRW ($a_{\rm{nea}}$), ARW and RR noise terms are read at $\tau$=1, 3, and $2^{0.5}$ respectively. The BI corresponds to the minimum at 0 slope. (b) $S(f)^{\scriptscriptstyle\frac{1}{2}}$ with $\Omega_{\rm{m}}=2\pi\times 133\,$Hz. The photodetector noise density $a_{\rm{det}}=1.9$\,\textmu g$\mathord{\cdot}$Hz$^{-\scriptscriptstyle\frac{1}{2}}$ is larger than the thermal limit of sensitivity, $a_{\rm{th}}=0.24\,$\textmu g$\mathord{\cdot}$Hz$^{-\scriptscriptstyle\frac{1}{2}}$. }
\label{allanbest}
\end{figure}

The MATLAB GUI AVAR \cite{ogier} is used to process an overlapped Allan deviation with 100 sample averaging period. The same GUI is used to fit and extract the noise co-efficient of each noise term, $q$, based on the slope of the plot defined by $\beta$. The velocity random walk (VRW, equivalent to the noise density, $a_{\rm{nea}}$) equals $q\tau^{-0.5}$ with $\tau=1\,$s. The bias instability (BI) equals $q\tau^{0}$, the acceleration random walk (ARW) equals $q\tau^{0.5}$ at $\tau=3$, and the rate ramp (RR) equals $q\tau^{1}$ at $\tau=2^{0.5}$. We refer the reader to \cite{niu} for more details about these inertial sensor errors. The black traces in Fig.~\ref{allanbest}\,(a)\,\&\,(b) show $\sigma$ and $S(f)^{\scriptscriptstyle\frac{1}{2}}$ for the WGM accelerometer studied here.

The VRW ($a_{\rm{nea}}$) is 4.5\,\textmu g$\mathord{\cdot}$Hz$^{-\scriptscriptstyle\frac{1}{2}}$, the BI is 31.8\,\textmu g, the ARW is 35.3\,\textmu g$\mathord{\cdot}$Hz$^{-\scriptscriptstyle\frac{1}{2}}$\,s$^{-1}$ and the RR is 11.7\,\textmu g$\mathord{\cdot}$s$^{-1}$. Of greatest importance for applications are VRW and BI where the BI indicates the minimum achievable DC reading limited by flicker noise and fluctuations to $d_{\mathrm{0}}$ or the optical field. ARW and RR are primarily related to temperature changes which can be reduced with stabilization. Section~\ref{longterm} explores the effect of PZT creep as a source of BI, ARW and RR noise. 

Comparing $a_{\rm{th}}$ and $a_{\rm{nea}}$ (VRW) shows that $a_{\rm{nea}}$ is over 18$\times$ larger. By analyzing the noise from the photodetector with the laser switched off (red trace) in Fig.~\ref{allanbest}\,(a), one finds $a_{\rm{det}}=1.9$\,\textmu g$\mathord{\cdot}$Hz$^{-\scriptscriptstyle\frac{1}{2}}$, indicating that detection noise, combined with additional laser noise, prohibits reaching $a_{\rm{th}}$.

Table~\ref{allanerror} summarizes the noise terms measured for the WGM accelerometer. Compared with the y-axis acceleromter of a commercial navigation device, the 3DM-GX3 IMU (made by LORD, MicroStrain Sensing Systems and characterized in \cite{ferrer}), the WGM accelerometer has a lower VRW and BI.  
\begin{table}[t!]
\renewcommand{\arraystretch}{1}
\caption{Noise parameters of the WGM accelerometer.}
\label{allanerror}
\centering
\begin{threeparttable}
\begin{tabular}{l c c c c}
\toprule
Noise (Units) & $\sigma(\tau)$ & Noise & WGM  & 3DM- \\
& gradient, $\beta$ & co-efficient, $q$ & sensor & GX3  \\ 
\midrule
VRW \big(\textmu g$\mathord{\cdot}$$\mathrm{Hz}^{-\scriptscriptstyle\frac{1}{2}}\big)$ & -0.5 & 4.5 & 4.5 & 76 \\
BI \big(\textmu g$\big)$  & 0 & 31.8 & 31.8 & 132\\
ARW \big(\textmu g$\mathord{\cdot}\mathrm{Hz}^{-\scriptscriptstyle\frac{1}{2}}\mathord{\cdot}\mathrm{s}^{-1}\big)$ & +0.5 & 20.4 & 35.3 & 12\\
RR \big(\textmu g$\mathord{\cdot}\mathrm{s}^{-1}\big)$ & +1 & 8.3 & 11.7 & -\\
\bottomrule
\end{tabular}
\begin{tablenotes}
\footnotesize
\item The noise parameters are velocity random walk (VRW=$a_{\rm{nea}}$), bias instability (BI), acceleration random walk (ARW) and rate ramp error (RR). A comparison with the y-axis accelerometer of the 3DM-GX3 IMU (LORD MicroStrain) is shown, measured in \cite{ferrer}.
\end{tablenotes} 
\end{threeparttable}
\end{table}
Our sensor has a similar VRW to a Fabry-Perot optomechanical accelerometer \cite{taylor2014} but with an order of magnitude lower BI. The current specification of the WGM accelerometer would allow for bio-tracking using ballistocardiography (BCG), whereby the BCG signal is generated at every heart beat as a recoil force from blood flowing into the aorta and distributed throughout the body. A BCG device is already sold by Murata using accelerometers with noise densities of 15\,\textmu g$\mathord{\cdot}$$\mathrm{Hz}^{-\scriptscriptstyle\frac{1}{2}}$ \cite{murata}. 

\subsection{Long Term Stability}
\label{longterm}
\begin{figure}[h!]
\centering
\includegraphics[width=8.8cm]{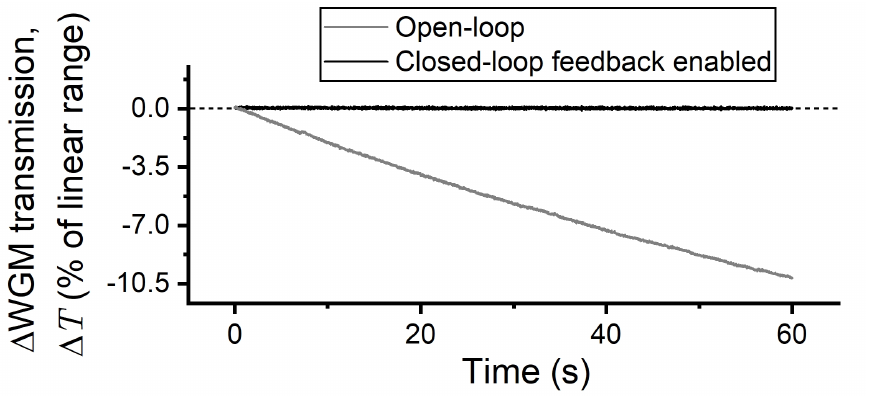}
\caption{\label{drift} Operating the accelerometer open-loop shows a slow drift due to PZT creep resulting in accumulation of a false acceleration in time (grey trace). By employing closed-loop feedback the drift is eliminated (black trace). 0\,g is indicated in dashes. }
\end{figure}

When used for applications such as navigation, accelerometers must have long-term stability, reducing the need for re-calibration. Due to the use of an open-loop PZT to translate $d_{\mathrm{0}}$ we observe a slow drift over time for time periods exceeding 10\,s due to known issues with PZT stability, creep and temperature effects \cite{gweon}, as shown in the grey trace of Fig.~\ref{drift}. This creates a false deflection which can result in substantial acceleration errors as well as altering the $d_{\mathrm{0}}$-dependent scale-factor investigated earlier in this paper. For long periods of operation creep can cause the microsphere to touch the tapered waveguide. Due to strong Van der Waals and electrostatic forces, it requires a large force to subsequently separate them, which is not practical to achieve in the field. Employing a closed-loop feedback scheme is therefore crucial to ensure $d_{\mathrm{0}}$ is kept stable. Feedback is also required for inertial test-masses with low $k$ which are susceptible to damage, especially shocks exceeding 100\,g if the sensor is dropped onto a hard surface. To this end, we send the WGM sensor output through a digital 100\,Hz low pass filter whose output is amplified and used as a proportional closed-loop feedback signal to the PZT supporting the microsphere-cantilever. This stabilizes $d_{\mathrm{0}}$ as shown in the black trace of Fig.~\ref{drift}. The feedback reduces the BI and RR by over a factor of 3 and 1000 respectively.

In fact, all commercial navigation grade inertial sensors employ sophisticated closed-loop feedback across the entire sensing bandwidth \cite{ragot}, with further error correction through the use of Kalman filtering \cite{ulbricht, hernandez}; a linear quadratic algorithm that analyzes a series of noisy measurements to produce estimates of position more accurately than a single measurement alone.

\section{Discussion}
The current limits on sensitivity due to classical laser noise and photodetector noise have been highlighted in the preceding section. Both can be significantly reduced, for example, by using shot-noise limited detection via a balanced homodyne method. This detection method has achieved displacement noise densities of $1\times 10^{-18}\,$m$\mathord{\cdot}$$\mathrm{Hz}^{-\scriptscriptstyle\frac{1}{2}}$ \cite{rousseau06, krause, schliesser18} where the latter reference is a WGM system with optomechanical rates similar to those found here. This would provide a 10$^{6}\times$ reduction of our noise floor, although a conservative aim is a factor of 1000 by using photodetectors with a noise equivalent power of $10^{-15}$\,W$\mathord{\cdot}$$\mathrm{Hz}^{-\scriptscriptstyle\frac{1}{2}}$ instead of the current value of $10^{-12}$\,W$\mathord{\cdot}$$\mathrm{Hz}^{-\scriptscriptstyle\frac{1}{2}}$.

\begin{figure}[h!]
\centering
\includegraphics[width=8.8cm]{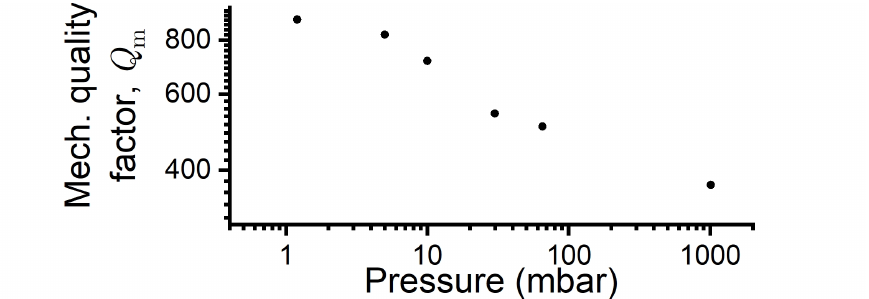}
\caption{\label{mechquality} The mechanical quality factor, $Q_{\mathrm{m}}$, of the silica microsphere-cantilevers can be increased by using lower environmental pressures. A $\times\,2.4$ improvement in $Q_{\rm{m}}$ is found at 1.2\,mbar.  }
\end{figure}

Assuming one continues to use hand-fabricated silica microsphere-cantilevers, increasing the mechanical quality factor $Q_{\rm{m}}$ by lowering the pressure will improve $a_{\rm{th}}$. An example of the change of $Q_{\rm{m}}$ with pressure is shown in Fig.~\ref{mechquality} where $Q_{\rm{m}}$ increases by a factor of $2.4$ at 1.2\,mbar. Lower pressures were not obtained. 

Considering these improvements, it is feasible to obtain $a_{\rm{nea}}<10\,$ng$\mathord{\cdot}$$\mathrm{Hz}^{-\scriptscriptstyle\frac{1}{2}}$, which would be sufficient to detect $a_{\rm{th}}=15\,$ng$\mathord{\cdot}$$\mathrm{Hz}^{-\scriptscriptstyle\frac{1}{2}}$ of the WGM accelerometer at 1\,mbar. Assuming a BI of 50\,ng, it would be possible to detect the 0.7\,\textmu g signal from a 50\,m radius sphere buried 100\,m underground with a density contrast of 2000\,kg$\mathord{\cdot}$$\mathrm{m}^{3}$. For unaided navigation, an error of $150\,$ng contributes a position error of 2.6\,mm in 60\,s. A lower pressure could be used to rival the current best noise density of $a_{\rm{nea}}=4\,$ng$\mathord{\cdot}$$\mathrm{Hz}^{-\scriptscriptstyle\frac{1}{2}}$ at 100\,Hz (Meggitt IEPE Model 86 accelerometer) \cite{endevco}. 

Moving to a MEMS manufacturing process where both waveguide and WGM test-mass are fabricated from a single bulk material, similar to commercial capacitive accelerometers utilizing interdigitated test-masses, will allow for significant size reduction and eliminate the use of a PZT for improved mechanical stability. Using a denser material such as silicon carbide for the microsphere-cantilever may improve the sensitivity by increasing $m_{\rm{eff}}$, whereas changing the cantilever length to increase $\Omega_{\rm{m}}$ extends the sensing bandwidth.

\section{Conclusion}
We have described and characterized the performance of a WGM accelerometer through modeling and testing. A sensor noise density of 4.5\,\textmu g$\mathord{\cdot}$$\mathrm{Hz}^{-\scriptscriptstyle\frac{1}{2}}$ and a bias instability of 31.8\,\textmu g is measured, rivaling the performance of commercial sensors for navigation as well as other optomechanical systems. Such a sensor is already suitable for non-contact bio-monitoring of heart rate variability as blood is accelerated through the aortic valves \cite{murata}. Detection of close-range gravitational forces such as the gravitational coupling between a cantilever test-mass and a driven 0.5\,mm diameter gold source-mass, as proposed in \cite{aspelmeyer}, is also feasible with our current system parameters.

For future development beyond the state-of-the-art, we have identified sources of noise that mask the detection sensitivity and contribute to long term drift, notably laser noise, photodetector noise, and noise from the piezostack actuator. We aim to reduce this background noise by a conservative factor of 1000 using a combination of active laser frequency and intensity stabilization combined with low noise photodetection. Operating the WGM accelerometer in a modest vacuum of 1\,mbar would allow for $a_{\rm{nea}}<a_{\rm{th}}<\,$20\,ng$\mathord{\cdot}$$\mathrm{Hz}^{-\scriptscriptstyle\frac{1}{2}}$. Although the Meggitt IEPE Model 86 accelerometer has $a_{\rm{nea}}=4\,$ng$\mathord{\cdot}$$\mathrm{Hz}^{-\scriptscriptstyle\frac{1}{2}}$, it lacks a response to DC forces. Our initial demonstration of closed-loop feedback also shows promising extension to wide bandwidth feedback that would allow for low stiffness cantilevers to be deployed in field-trials as well as within the lab. With a MEMS fabricated microsphere-cantilever, electrostatic \cite{miao12} or optical gradient forces \cite{painter, gavartin12} can be used instead of piezoelectric actuation which would further minimize sources of noise. Such a chip-scale sensor could be deployed in CubeSats for detection of gravitational waves.

\ifCLASSOPTIONcaptionsoff
  \newpage
\fi

\bibliographystyle{IEEEtran}

\bibliography{IEEEabrv,Li_bib}

\begin{thebibliography}{10}
\providecommand{\url}[1]{#1}
\csname url@samestyle\endcsname
\providecommand{\newblock}{\relax}
\providecommand{\bibinfo}[2]{#2}
\providecommand{\BIBentrySTDinterwordspacing}{\spaceskip=0pt\relax}
\providecommand{\BIBentryALTinterwordstretchfactor}{4}
\providecommand{\BIBentryALTinterwordspacing}{\spaceskip=\fontdimen2\font plus
\BIBentryALTinterwordstretchfactor\fontdimen3\font minus
  \fontdimen4\font\relax}
\providecommand{\BIBforeignlanguage}[2]{{%
\expandafter\ifx\csname l@#1\endcsname\relax
\typeout{** WARNING: IEEEtran.bst: No hyphenation pattern has been}%
\typeout{** loaded for the language `#1'. Using the pattern for}%
\typeout{** the default language instead.}%
\else
\language=\csname l@#1\endcsname
\fi
#2}}
\providecommand{\BIBdecl}{\relax}
\BIBdecl

\bibitem{schliesser18}
A.~Schliesser, O.~Arcizet, R.~Rivi{\'e}re, G.~Anetsberger, and T.~J.
  Kippenberg, ``Resolved-sideband cooling and position measurement of a
  micromechanical oscillator close to the {H}eisenberg uncertainty limit,''
  \emph{Nature Physics}, vol.~5, pp. 509--514, Jun. 2009.

\bibitem{taylor2014}
O.~Gerberding, F.~G. Cervantes, J.~Melcher, J.~R. Pratt, and J.~M. Taylor,
  ``Optomechanical reference accelerometer,'' \emph{Metrologia}, vol.~52,
  no.~5, pp. 654--665, Sep. 2015.

\bibitem{lia}
Y.~L. Li, J.~Millen, and P.~F. Barker, ``Simultaneous cooling of coupled
  mechanical oscillators using whispering gallery mode resonances,''
  \emph{Optics Express}, vol.~24, no.~2, pp. 1392--1401, Jan. 2016.

\bibitem{pernice2009}
M.~Li, W.~H.~P. Pernice, and H.~X. Tang, ``Reactive cavity optical force on
  microdisk-coupled nanomechanical beam waveguides,'' \emph{Phys. Rev. Lett.},
  vol. 103, p. 223901, Nov. 2009.

\bibitem{painter}
M.~Eichenfield, C.~P. Michael, R.~Perahia, and O.~Painter, ``Actuation of
  micro-optomechanical systems via cavity-enhanced optical dipole forces,''
  \emph{Nature Photonics}, vol.~1, pp. 416--422, Jul. 2007.

\bibitem{transduction}
R.~Madugani, Y.~Yang, J.~M. Ward, V.~H. Le, and S.~N. Chormaic,
  ``Optomechanical transduction and characterization of a silica microsphere
  pendulum via evanescent light,'' \emph{Appl. Phys. Lett.}, vol. 106, p.
  241101, Dec. 2015.

\bibitem{miao12}
H.~Miao, K.~Srinivasan, and V.~Aksyuk, ``A microelectromechanically controlled
  cavity optomechanical sensing system,'' \emph{New Journal of Physics},
  vol.~14, p. 075015, Jul. 2012.

\bibitem{gavartin12}
E.~Gavartin, P.~Verlot, and T.~J. Kippenberg, ``A hybrid on-chip optomechanical
  transducer for ultrasensitive force measurements,'' \emph{Nature
  Nanotechnology}, vol.~7, pp. 509--514, Jun. 2012.

\bibitem{kippenbergg}
G.~Anetsberger, E.~M. Weig, J.~P. Kotthaus, and T.~J. Kippenberg, ``Cavity
  optomechanics and cooling nanomechanical oscillators using microresonator
  enhanced evanescent near-field coupling,'' \emph{Comptes Rendus Physique},
  vol.~12, no. 9-10, pp. 800--816, Dec. 2011.

\bibitem{haus}
J.-P. Laine, C.~Tapalian, B.~Little, and H.~Haus, ``Acceleration sensor based
  on high--q optical microsphere resonator and pedestal antiresonant reflecting
  waveguide coupler,'' \emph{Sensors and Actuators A}, vol.~93, no.~1, pp.
  1--7, Aug. 2001.

\bibitem{barclay}
M.~Wu, A.~C. Hryciw, C.~Healey, D.~P. Lake, H.~Jayajumar, M.~R. Freeman, J.~P.
  Davis, and P.~E. Barclay, ``Dissipative and dispersive optomechanics in a
  nanocavity torque sensor,'' \emph{Phys. Rev. X}, vol.~4, p. 021052, Jun.
  2014.

\bibitem{krause}
A.~G. Krause, M.~Winger, T.~D. Blasius, Q.~Lin, and O.~Painter, ``A
  high-resolution microchip optomechanical accelerometer,'' \emph{Nature
  Photonics}, vol.~6, pp. 768--772, Mar. 2012.

\bibitem{ligo2016}
{LIGO Scientific Collaboration \& Virgo Collaboration}, ``Observation of
  gravitational waves from a binary black hole merger,'' \emph{Phys. Rev.
  Lett.}, vol. 116, p. 061102, Feb. 2016.

\bibitem{rousseau06}
O.~Arcizet, P.~F. Cohadon, T.~Briant, M.~Pinard, A.~Heidmann, J.~M. Mackowski,
  C.~Michel, L.~Pinard, O.~Fran\c{c}ais, and L.~Rousseau, ``High sensitivity
  optical monitoring of a micromechanical resonator with a quantum limited
  optomechanical sensor,'' \emph{Phys. Rev. Lett.}, vol.~97, p. 133601, Sep.
  2006.

\bibitem{ulbricht}
A.~Setter, M.~Toro{\v s}, J.~F. Ralph, and H.~Ulbricht, ``Real-time {K}alman
  filter: Cooling of an optically levitated nanoparticle,'' \emph{Phys. Rev.
  A.}, vol.~97, p. 033822, Mar. 2018.

\bibitem{natale}
G.~Gagliardi, M.~Salza, D.~Avino, P.~Ferraro, and P.~D. Natale, ``Probing the
  ultimate limit of fiber-optic strain sensing,'' \emph{Science}, vol.~19, pp.
  1081--1084, Nov. 2010.

\bibitem{pajak1998}
J.~Kalenik and R.~Pajak, ``A cantilever optical-fiber accelerometer,''
  \emph{Sensors and Actuators A: Physical}, vol.~68, no. 1-3, pp. 350--355,
  Jun. 1998.

\bibitem{ferrer}
A.~G. Quinchia, G.~Falco, E.~Falletti, F.~Dovis, and C.~Ferrer, ``A comparison
  between different error modeling of {MEMS} applied to {GPS/INS} integrated
  systems,'' \emph{Sensors}, vol.~13, no.~8, pp. 9549--9588, Jul. 2013.

\bibitem{inertiallabs}
``High performance advanced {MEMS} industrial \& tactical grade inertial
  measurement units, {IMU-P} {R}ev. 2.0,'' Inertial Labs, Virginia, USA.

\bibitem{vahala}
M.~Cai, O.~Painter, and K.~J. Vahala, ``Observation of critical coupling in a
  fiber taper to a silica-microsphere whispering-gallery mode system,''
  \emph{Phys. Rev. Lett.}, vol.~85, no.~74, pp. 74--77, Jul. 2000.

\bibitem{haus1}
H.~A. Haus, \emph{Waves and Fields in Optoelectronics}.\hskip 1em plus 0.5em
  minus 0.4em\relax Englewood Cliffs, NJ: Prentice-Hall, 1984.

\bibitem{carmon}
T.~Carmon, L.~Yang, and K.~J. Vahala, ``Dynamical thermal behaviour and thermal
  self-stability of microcavities,'' \emph{Optics Express}, vol.~12, no.~20,
  pp. 4742--4750, Sep. 2004.

\bibitem{ogier}
\BIBentryALTinterwordspacing
E.~Ogier. (2016, Nov.) Avar. MATLAB Central File Exchange. [Online]. Available:
  \url{https://uk.mathworks.com/matlabcentral/fileexchange/55765-avar}
\BIBentrySTDinterwordspacing

\bibitem{niu}
N.~El-Sheimy, H.~Hou, and X.~Niu, ``Analysis and modeling of inertial sensors
  using allan variance,'' \emph{{IEEE} Transactions on Instrumentation and
  Measurement}, vol.~57, no.~1, pp. 140--149, Jan. 2008.

\bibitem{murata}
U.~Merihein{\"a}, ``{BCG} measurements in beds,'' Murata, Whitepaper 8375,
  2017.

\bibitem{gweon}
H.~Jung and D.-G. Gweon, ``Creep characteristics of piezoelectric actuators,''
  \emph{Rev. of Sci. Instrum.}, vol.~71, no.~4, pp. 1896--1900, Apr. 2000.

\bibitem{ragot}
P.~Zwahlen, Y.~Dong, A.-M. Nguyen, F.~Rudolf, J.-M. Stauffer, P.~Ullah, and
  V.~Ragot, ``Breakthrough in high performance inertial navigation grade
  sigma-delta {MEMS} accelerometer,'' in \emph{Proc. {IEEE}/ION Position,
  Location and Navigation Symposium 2012}, Myrtle Beach, USA, Apr. 2012, pp.
  15--19.

\bibitem{hernandez}
W.~Hern{\'a}ndez, ``Improving the responses of several accelerometers used in a
  car under performance tests by using kalman filtering,'' \emph{Sensors},
  vol.~1, no.~1, pp. 38--52, Jun. 2001.

\bibitem{endevco}
F.~A. Levinzon, ``Ultra-low-noise seismic piezoelectric accelerometer with
  integral {FET} amplifier,'' \emph{{IEEE} Sensors Journal}, vol.~12, no.~6,
  pp. 2262--2268, Feb. 2012.

\bibitem{aspelmeyer}
J.~Schm{\"o}le, M.~Dragosits, H.~Hepack, and M.~Aspelmeyer, ``A micromechanical
  proof-of-principle experiment for measuring the gravitational force of
  milligram masses,'' \emph{Class. Quantum Grav.}, vol.~33, no.~4, p. 125031,
  May 2016.

\end{thebibliography}

\end{document}